\documentclass[twocolumn]{aastex631}
\usepackage{array, makecell}
\usepackage{multirow}
\usepackage{comment}

\newcommand{\as}{$^{\prime\prime}$}

\begin{document}

\title{Hinode EIS Observations of Plasma Composition Evolution \\ and Radiative Cooling of Solar Flare Loops}

\correspondingauthor{Teodora Mihailescu}
\email{teodora.mihailescu@inaf.it}

\author[0000-0001-8055-0472]{Teodora Mihailescu}
\altaffiliation{Current institution: INAF Osservatorio Astronomico di Roma, 00078 Monte Porzio Catone, Italy}
\affil{NASA Goddard Space Flight Center, Greenbelt, MD 20771, USA}
\affil{Universities Space Research Association, Washington, DC 20024, USA}
\affil{Mullard Space Science Laboratory, University College London, Surrey, RH5 6NT, UK}

\author[0000-0001-9034-2925]{Peter R. Young}
\affil{NASA Goddard Space Flight Center, Greenbelt, MD 20771, USA}
\affil{Northumbria University, Newcastle upon Tyne, NE1 8ST, UK}

\author[0000-0002-2189-9313]{David H. Brooks}
\affil{Computational Physics Inc., Springfield, VA, 22151, USA}
\affil{Mullard Space Science Laboratory, University College London, Surrey, RH5 6NT, UK}

\author[0000-0002-0665-2355]{Deborah Baker}
\affil{Mullard Space Science Laboratory, University College London, Surrey, RH5 6NT, UK}

\author[0000-0002-0053-4876]{Lucie M. Green}
\affil{Mullard Space Science Laboratory, University College London, Surrey, RH5 6NT, UK}

\author[0000-0003-3137-0277]{David M. Long}
\affil{Dublin City University, Dublin, D09V209, Ireland}
\affil{Dublin Institute for Advanced Studies, Dublin D15XR2R, Ireland}

\author[0000-0002-2943-5978]{Lidia van Driel-Gesztelyi}
\affil{Mullard Space Science Laboratory, University College London, Surrey, RH5 6NT, UK}
\affil{Konkoly Observatory, Konkoly Thege út 15-17., H-1121, Budapest, Hungary}



\begin{abstract}
Plasma composition in flaring regions has been shown to have significant spatial and temporal variations, likely driven  by dynamical processes that take place as a consequence of the sudden energy release at the reconnection site. The origins of these variations, as well as the effects they might, in turn, have on flare loops dynamics are not yet fully understood.
In this work, we investigate the link between flare loop cooling times and plasma composition evolution in the loops formed during the M-class flare peaking at 13:56 UT on the 2022 April 2 using high cadence Hinode EIS spectroscopic observations. 
The analysis focuses on quantifying the cooling rate (using a series of emission lines covering a wide temperature range) and plasma composition evolution (using the Ca {\scriptsize XIV} 193.866 \AA/Ar {\scriptsize XIV} 194.401 \AA\ diagnostic) at the apex and footpoint of the flare loop arcade. Results show slower cooling and a FIP bias of $2.4\pm0.2$ in the loop footpoint and faster cooling and a stronger FIP bias of $2.8 \pm 0.2$ in the loop apex.
The potential effects of plasma composition changes on the radiative cooling process of flare loops are also investigated by comparing observed loop cooling times to those predicted by simulations from the EBTEL 0D hydrodynamic model. The EBTEL simulations show that an higher FIP bias would lead to a faster radiative cooling rate and, therefore, shorter cooling times. This suggests that the variation in FIP bias observed in the two features could be responsible for the different cooling times observed.
\end{abstract}



\section{Introduction} \label{Introduction}
The plasma composition of the solar corona is known to vary spatially and in time, in spite of the photosphere underneath having constant and homogeneous composition \citep[e.g.][]{asplund_chemical_2009, asplund_chemical_2021}. The variation is observed as an enhancement in the relative abundance of elements with low first ionization potential (FIP; $<10$ eV) compared to those with a high FIP ($>10 $ eV). This is called the FIP effect. The strength of the FIP effect is quantified by the FIP bias parameter. For a given element X, the FIP bias is defined as:
\begin{equation}    \mathrm{FIP}_\mathrm{bias}=\frac{\mathrm{A}_\mathrm{X, corona}}{\mathrm{A}_\mathrm{X, photosphere}},
\end{equation}
where $\mathrm{A}_\mathrm{X, corona}$ and $\mathrm{A}_\mathrm{X, photosphere}$ represent the abundance of element X relative to hydrogen in the corona and the photosphere respectively. The process driving the FIP effect is believed to take place in the chromosphere and manifest as a preferential acceleration of low-FIP elements to the transition region \citep{laming_unified_2004, laming_fip_2015, reville_investigating_2021, martinez-sykora_impact_2023}. From the transition region, plasma flows bring the plasma with modified abundance into the corona where the effect can be observed with current instrumentation. The properties of this flow (speed, mass flux) determine how fast changes in plasma composition at transition region level are reflected in the corona. In addition, once in the corona, plasma composition can also be changed as a result of plasma mixing. This can happen when, for example, magnetic reconnection between two loops with different composition mixes the plasma along those loops leading to an averaged composition in the two post-reconnection loops \citep[e.g.,][]{baker_fip_2015, baker_evolution_2022}. This is independent of the FIP effect itself.

The plasma composition in flare loops appears to be different from that of quiescent active region loops. Unlike quiescent active region loops, where fractionated plasma is supplied to the corona via the loop footpoints \citep{baker_plasma_2013}, in flare loops, the corona is supplied with unfractionated plasma from the lower atmosphere. Results from full-Sun spectra in the EUV \citep{warren_measurements_2014} and X-rays \citep[e.g.][]{mondal_evolution_2021, nama_coronal_2023} found evidence that the average FIP bias, dominated by the flare emission, evolves from quiet Sun values in the preflare phase to photospheric values around the peak of the GOES soft X-ray flux, and returns to its original value over the course of a few minutes to a few hours \citep[significantly faster than the timescales of hours to days observed in quiescent active regions, e.g.,][]{widing_rate_2001, baker_fip_2015, ko_correlation_2016}. This evolution of the FIP bias seems logical in the context of chromospheric evaporation in flares \citep{warren_measurements_2014}, whereby plasma is rapidly expelled from the chromosphere via evaporation at speeds of up to 200 km s$^{-1}$ \citep[e.g.][]{milligan_velocity_2009, brosius_chromospheric_2013}, i.e., significantly faster than the steady upflow speeds of a few km s$^{-1}$ typical in quiescent conditions \citep[e.g.][]{imada_self-organization_2012}. This means that either the plasma is expelled from the chromosphere before the processes driving the FIP effect can occur, hence its photospheric composition, or that it is expelled from below the region where the processes driving the FIP effect takes place in the chromosphere. 

Spatially resolved observations from Hinode EIS show that the brighter flare loop apex has an increased FIP bias \citep[measured using the Ca {\scriptsize XIV} 193.866 \AA/Ar {\scriptsize XIV} 194.401 \AA\ diagnostic;][]{doschek_photospheric_2018}, and the FIP bias values decrease along the loop as the footpoint is approached \citep{doschek_photospheric_2018}, consistent with the scenario where plasma with photospheric composition is supplied via the footpoints. \citet{to_spatially_2024} proposed that this gradient is due to a combination of two different processes: chromospheric evaporation upflows supplying plasma with photospheric composition from lower altitudes and reconnection downflows from the plasma sheet confined to loop tops supplying the flare loop apex with plasma with high FIP bias. The reconnection downflow is expected to bring plasma with high FIP bias because the reconnecting loops are active region loops which typically show high FIP bias values.

The plasma composition pattern and evolution observed in solar flares can be a valuable tracer of mass and energy and, therefore, has the potential to provide insight into flare dynamics. For example, the composition of the plasma transported upwards via chromospheric evaporation during a flare must be linked to the amount of energy released by the flare and the chromospheric depth at which it is deposited. An extreme example of this is the inverse FIP effect (IFIP) occasionally observed at flare loop footpoints \citep{doschek_anomalous_2015, doschek_mysterious_2016, doschek_sunspots_2017}. In the IFIP case, low-FIP elements are under-abundant rather than over-abundant in the corona compared to high-FIP elements. According to the ponderomotive force model \citep{laming_fip_2015}, IFIP is driven low down in the chromosphere so an observation of IFIP suggests that the evaporation height must have been very low to be able to bring this plasma into the corona. This suggests that a large amount of energy was released. In addition, the plasma composition, in turn, can influence flare dynamics as it affects the radiative cooling process of loops, particularly in the temperature range $\log(T/\mathrm{K})=5.3-7.0$ \citep{cook_effect_1989}. \citet{brooks_diagnostic_2018} studied the flare loop cooling times associated with an IFIP producing solar flare to address the question of whether IFIP is produced by a depletion in low-FIP elements or an enhancement in high-FIP elements. They concluded that the first option is more likely in their event, as the second one would lead to much shorter cooling times than those observed.

Gaining better understanding of the plasma composition behaviour in flares requires spectroscopic observations that provide both spatial information and high cadence. Typically, observations that provide high cadence are full-Sun integrated, while those that provide spatial information have relatively low cadence (a few minutes). In the work presented here, we partially overcome this challenge by analysing sit-and-stare spectroscopic observations that captured the flare loop evolution in two key locations, the loop apex and footpoint, with high cadence. We also explore links between the plasma composition evolution and the cooling process of the flare loops following the method of \citet{brooks_diagnostic_2018}.

\section{Flare Loops  Evolution}
\begin{figure}
    \centering
    \includegraphics[width=0.5\textwidth]{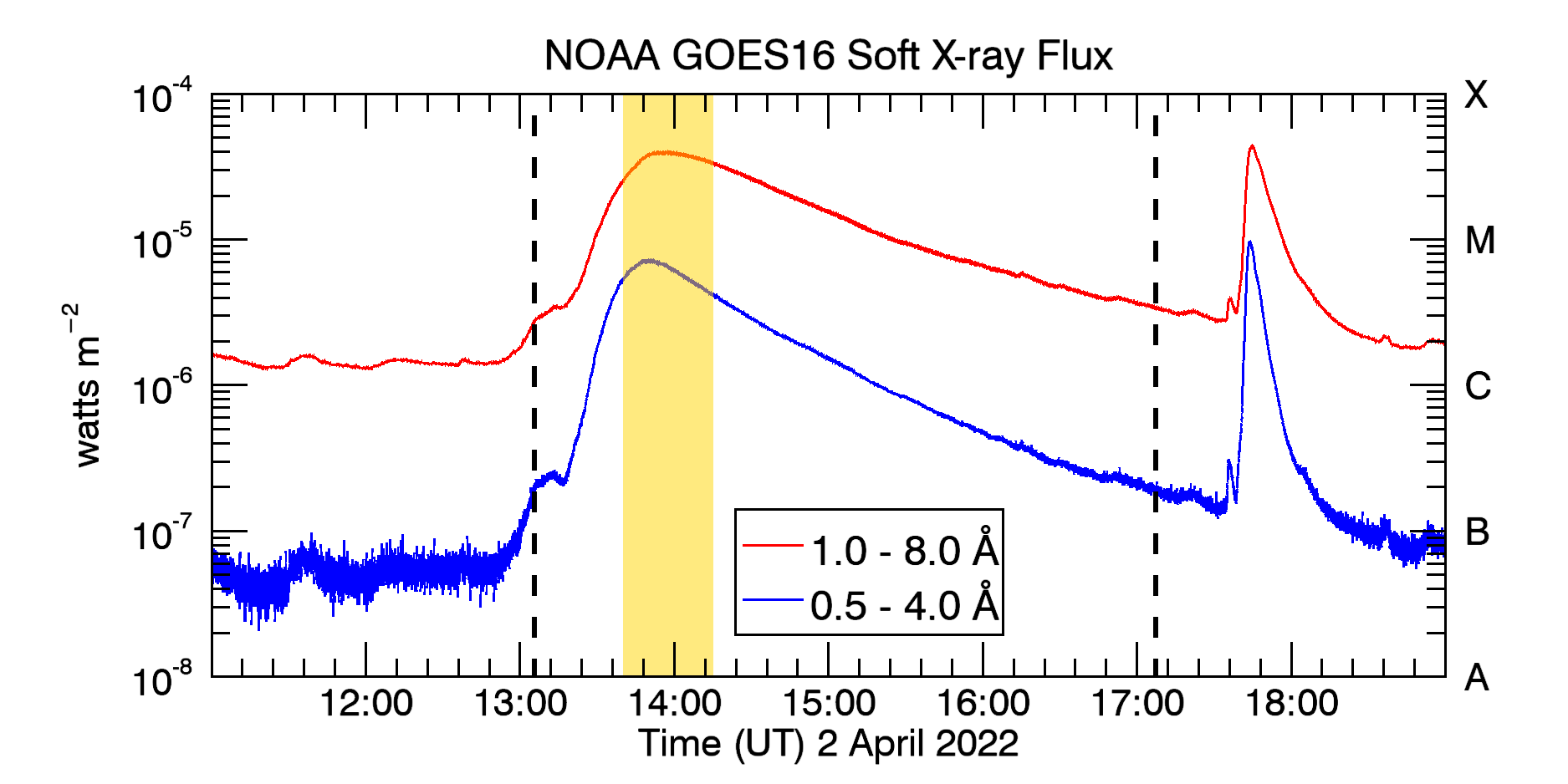}
    \caption{Soft X-ray flux observed in the two GOES channels (1-8 \AA\ and 0.5-4 \AA). The vertical dashed lines indicate the start and end times of the Hinode EIS observations, and the yellow rectangle shows the portion of the observations that this study focuses on.}
    \label{Lightcurve}
\end{figure}

\begin{figure}
    \centering
    \includegraphics[width=0.5\textwidth]{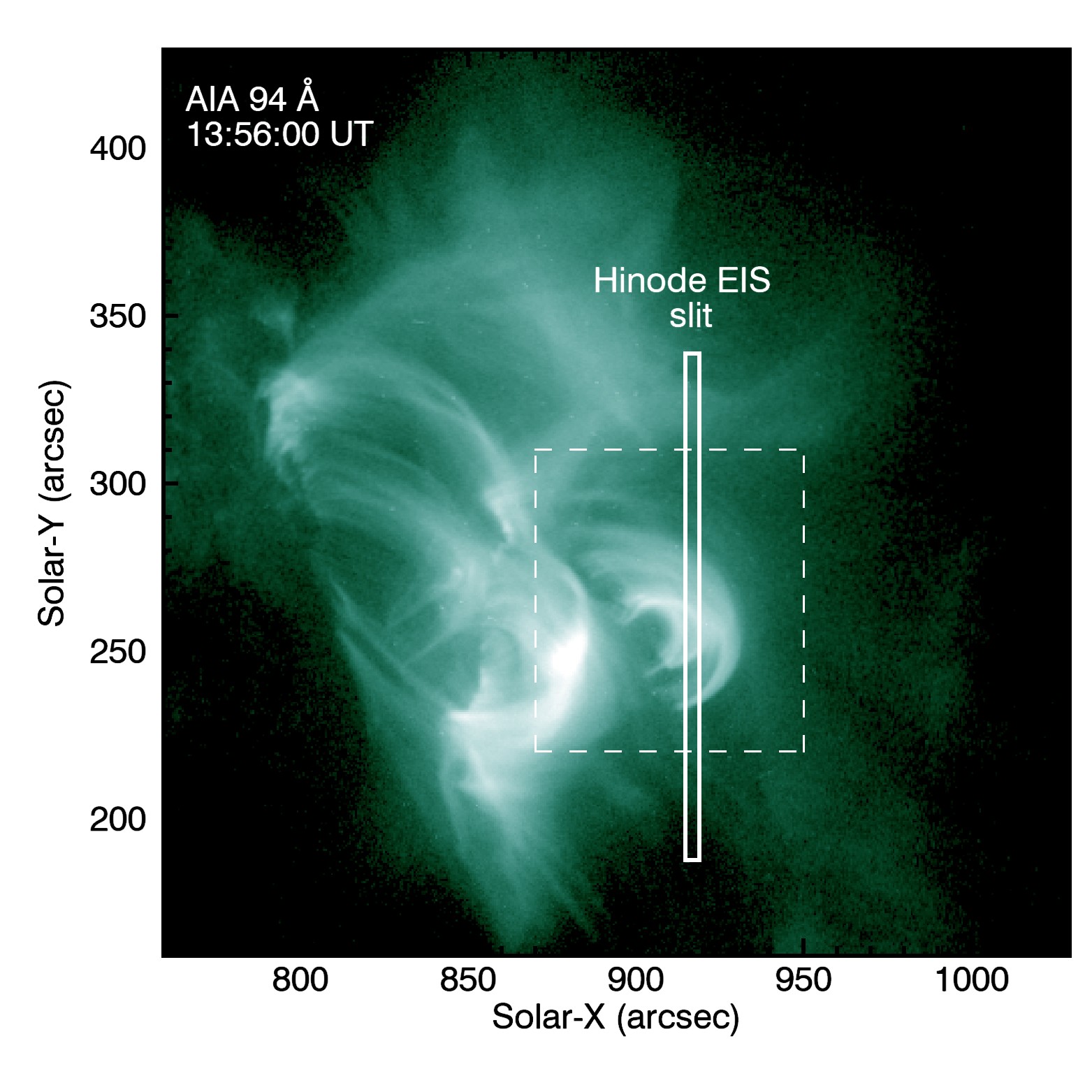}
    \caption{Context AIA 94 \AA\ image of the flare loops associated with the M3.9 flare peaking at  13:56 UT on 2022 April 2. The solid white rectangle indicates the position of the EIS slit (for Fe {\scriptsize XII} 195.119 \AA), and the dashed white rectangle shows the loop arcade studied in this work (see Figure \ref{AIApanel} for the evolution of this loop arcade). A movie version of this figure, covering the entire duration of the flare from 13:00 to 17:00 UT, is in the online Journal. The image has logarithmic scaling.}
    \label{AIAcontext}
\end{figure}

The flare under study is an M3.9 long duration event that started at approximately 13:00 UT and peaked at 13:56 UT on 2022 April 2 (see Figure \ref{Lightcurve}). It accompanied the eruption of a filament that had formed in between two large active regions, NOAA AR 12976 and NOAA AR 12975, and was destabilised by the emergence of a third active region, NOAA AR 12977 \citep[see][ for details on the eruption, including magnetic configuration and orientation of the polarity inversion lines]{janvier_multiple_2023}. It is the first of two M-class flares produced by this complex on this day, followed by a short duration M4.3 flare at 17:34 that EIS did not capture (see Figure \ref{Lightcurve}). The focus of this work is on the small flare loop arcade created by the M3.9 event, which is located around x=880--940\as\ and y=230--300\as\ (see Figure \ref{AIAcontext}) and was observed by Hinode EIS throughout the duration of the flare. These flare loops start brightening in the AIA images at approximately 13:40 UT and disappear completely by 14:30 UT (see Supplementary Movie 1 and Figure \ref{AIApanel}). 

\begin{figure*}
    \centering
    \includegraphics[width=1.0\textwidth]{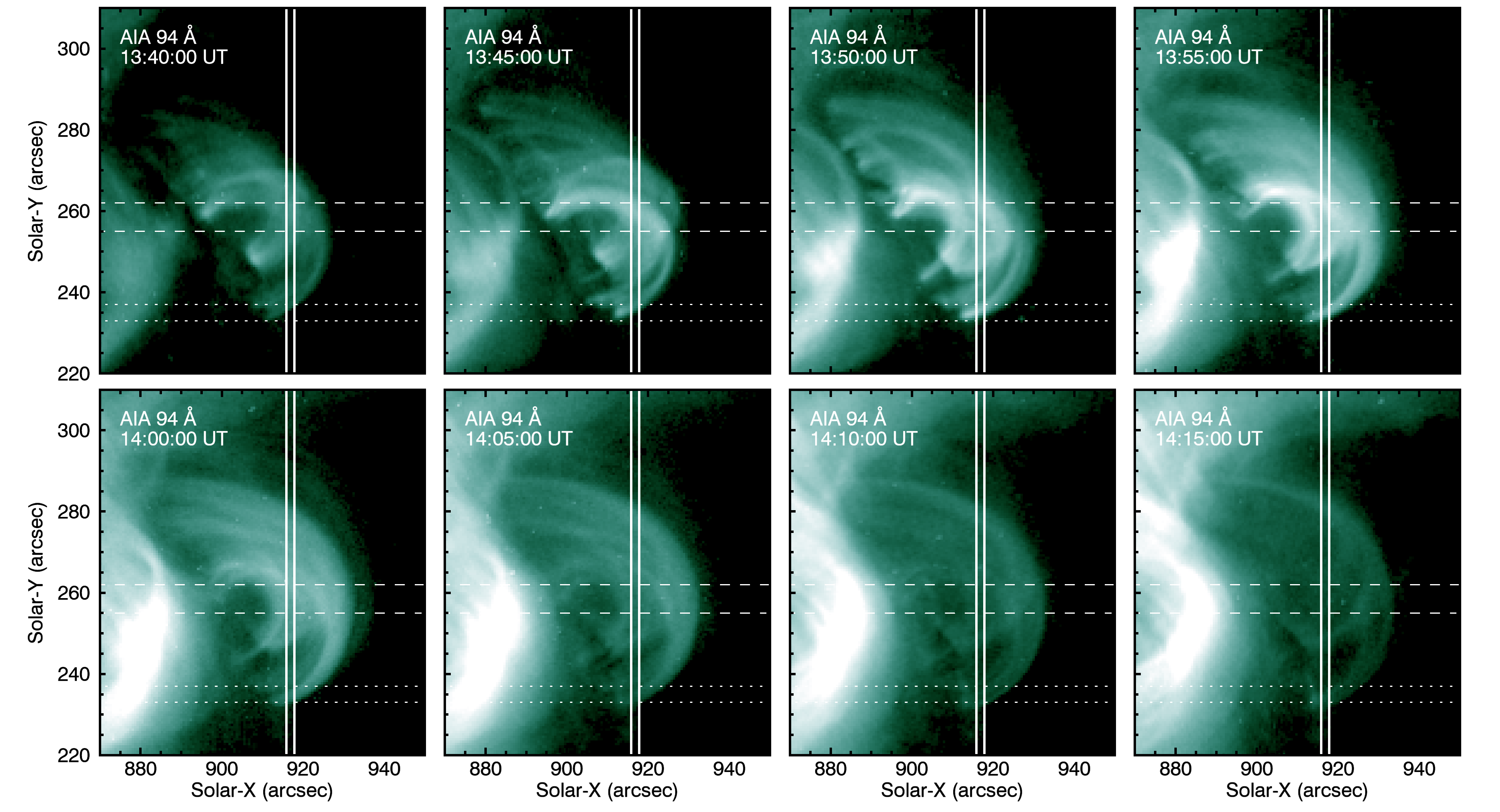}
    \caption{Close up evolution of the flare loop arcade observed by Hinode EIS, corresponding to the field of view shown in the dashed white rectangle in Figure \ref{AIAcontext}. The vertical solid white lines indicate the location of the EIS slit. The horizontal dashed and dotted lines at y=255\as--262\as{} and y=233\as--237\as{} indicate the regions selected for the loop top and loop footpoint analyses respectively. The image has logarithmic scaling.}
    \label{AIApanel}
\end{figure*}

\section{Hinode EIS Observations}
The evolution of the flare loops was captured by the Hinode EUV Imaging Spectrometer \citep[EIS;][]{culhane_euv_2007} using a sit-and-stare observing study. The observation, consisting of 4 consecutive runs of the Flare\_SNS\_v2 study\footnote{Details at: \href{https://solarb.mssl.ucl.ac.uk/SolarB/estudylist.jsp}{https://solarb.mssl.ucl.ac.uk/SolarB/estudylist.jsp}}, started just before the beginning of the impulsive phase of the flare and continued through most of the decay phase (see Figure \ref{Lightcurve}). In this study, the EIS slit has a fixed position, which is only adjusted slightly to account for solar rotation. The slit was ideally located to capture the flare loops cooling during the decay phase with a cadence of 16.2 s (see Figures \ref{AIAcontext}, \ref{AIApanel}).

The present analysis focuses on the EIS observations taken around the peak of the flare, i.e., from approximately 13:40 UT to 14:15 UT (see yellow rectangle in Figure \ref{Lightcurve}). The level-0 data were obtained from the Hinode EIS archive\footnote{\href{https://solarb.mssl.ucl.ac.uk/SolarB/SearchArchive.jsp}{https://solarb.mssl.ucl.ac.uk/SolarB/SearchArchive.jsp}} using the IDL Solar Software library. The data were calibrated with the eis\_prep routine \citep[see EIS Software Note No. 1; ][]{young_calibrating_2025}, which removes the CCD dark current, cosmic ray pattern, and hot and dusty pixels from detector exposures and provides the radiometric calibration to physical units ($\text{erg } \text{cm}^{-2}\text{s}^{-1}\text{sr}^{-2}\text{\AA{}}^{-1}$). Each line used in this study (see Table \ref{Cooling_results}) was fitted with a single Gaussian using the eis\_auto\_fit routine (see EIS Software Note No. 16; \citealt{young_eis_auto_fit_2022}, and EIS Software Note No. 17; \citealt{young_eis_2020}). The most recent EIS calibration method \citep{del_zanna_hinode_2025} was used to calculate line intensities.

\begin{deluxetable}{ll}[t]
\centering
\tablecolumns{2}
\tablehead{\colhead{EIS Study Details}}
\tablecaption{Hinode EIS study details. \label{EIS_studies}}
\startdata
Raster times & \makecell[l]{02/04/2022 13:05 \\ 02/04/2022 14:06 \\ 02/04/2022 15:06 \\ 02/04/2022 16:07}\\
Study number & 555 \\
Raster type & sit-and-stare (SNS) \\
Window height & 152\as\\
Rastering & 2\as\ slit, 225 sets \\
Exposure time & 10 s \\
Exposure delay & 5 s \\
Readpout time & 1.2 s\\
Total cadence & 16.2 s\\
Total raster time & 1 h 0 m 20 s\\
\enddata
\tablecomments{More details available on the \href{https://solarb.mssl.ucl.ac.uk/SolarB/estudylist.jsp}{Hinode EIS studies page}}
\end{deluxetable}

To align the EIS rasters to AIA, the standard offset was obtained from the eis\_aia\_offsets routine and an additional manual correction was performed by comparing the EIS Fe {\scriptsize XII} 195.119 \AA\ intensity map with images from the AIA 193 \AA\ passband. For the additional correction, synthetic AIA rasters were produced from a sequence of images in the 193 \AA\ passband using the IDL Solarsoft aia\_make\_eis\_raster routine \citep[see EIS Software Note No. 26;][]{young_creating_2023}. An offset correction of (+8\as, +15\as), in addition to the default offset given by the eis\_aia\_offset routine, was found to give the best alignment. 

The analysis presented in this section was carried out using the CHIANTI Atomic Database Version 11.01 \citep{dere_chianti_1997, dufresne_chiantiatomic_2024}. 

\subsection{Flare Loop Cooling}
\label{S: Flare Loop Cooling}
The cooling flare loops are observed by Hinode EIS in a series of emission lines that form at a wide range of temperatures. Here, we focus on two key features captured by the EIS slit: the loop top located at y=255--262\as\ and the flare loop footpoint located at y=233-- 237\as\ (see Figure \ref{AIApanel}). As it is clear from Figure \ref{AIApanel}, these two features are part of different loops within the observed loop arcade: the apex is in a lower loop in the arcade than the footpoint. In fact, it is also possible that EIS is not observing the exact same feature in these two locations over the duration of the observation. What is referred to here as loop top or loop footpoint could actually be a series of snapshots of consecutive loop strands that reconnect and move through the EIS field of view over the duration of the observation. However, this flare loop arcade is made out of multiple loops that go through the same process of being formed and heated via reconnection, and then cooling down and shrinking. We, therefore, expect loops within the same loop arcade to have similar properties and cooling times. In this work, we treat the results obtained at the loop top and footpoint identified in Figure \ref{AIApanel} as representative for the loop tops and footpoints within the arcade. Note that, by footpoint, we refer to a lower portion of the coronal loop that is closer to the solar surface, and not a chromospheric footpoint.

\begin{figure}
    \centering
    \includegraphics[width=0.46\textwidth]{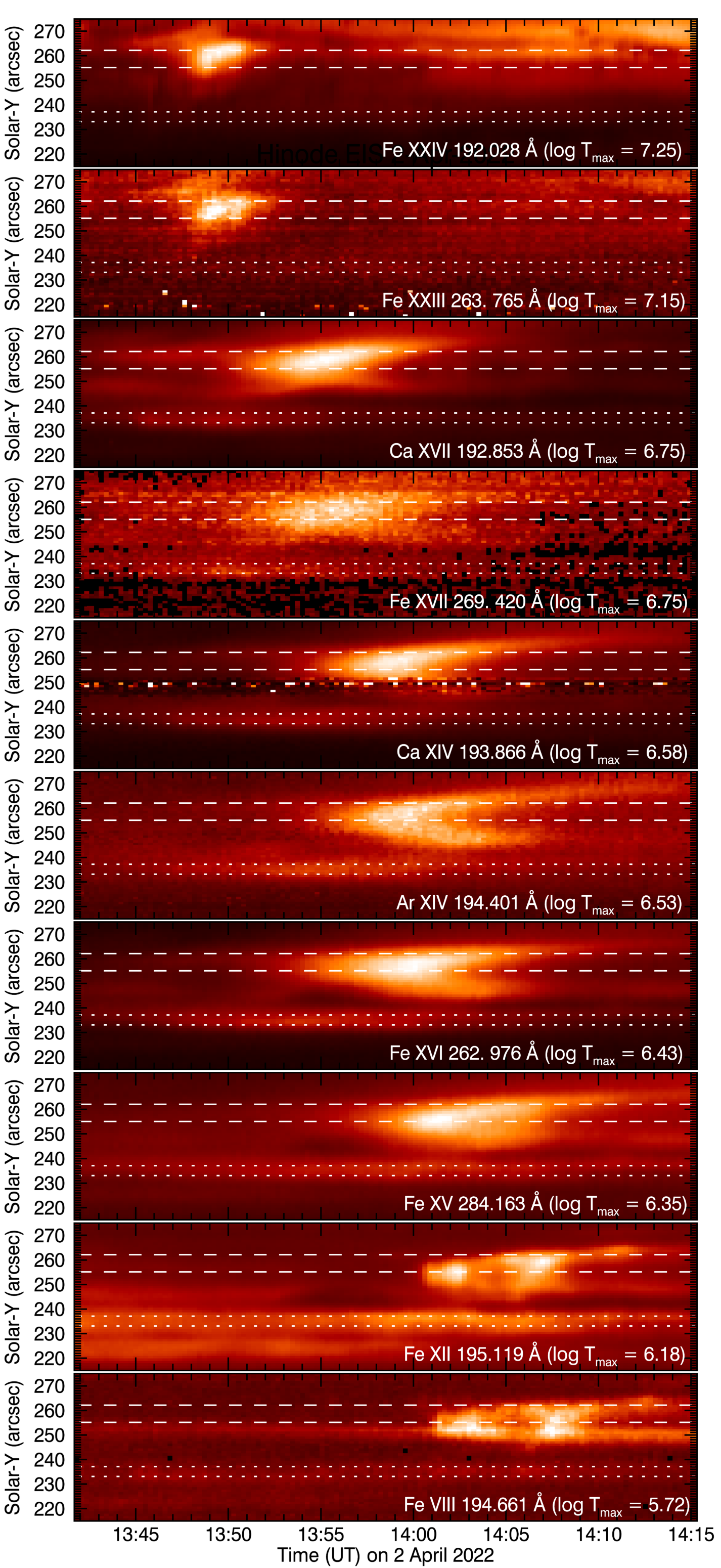}
    \caption{Flare loop emission evolution in progressively cooler lines from top to bottom. The dashed and dotted lines at y=255\as--262\as{} and y=233\as--237\as{} indicate the regions selected for the loop top and loop footpoint analyses, respectively (see Figures \ref{EIS intensity lineouts} and \ref{Plasma parameters results}). The image has linear scaling. The wavelengths and peak formation temperatures for each line are shown in the bottom right hand side of each panel.}
    \label{EISintensity}
\end{figure}

\begin{figure}
    \centering
    \includegraphics[width=0.46\textwidth]{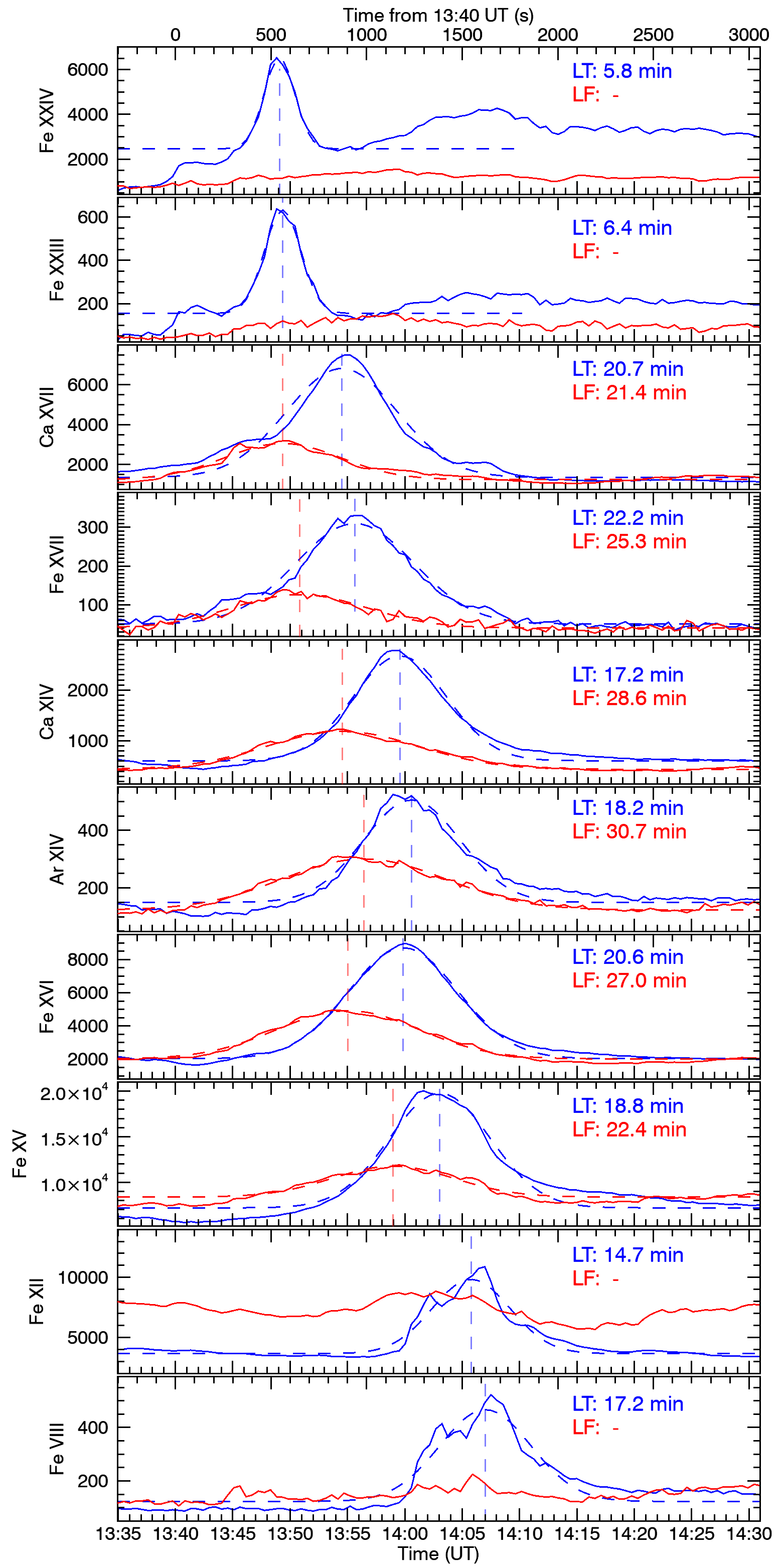}
    \caption{Intensity evolution in the loop top (blue, averaged between y=255\as--262\as) and footpoint (red, averaged between y=233\as--237\as). The Gaussian fits to the intensity evolution in each line and for each feature are shown in dashed lines. The peak emission times for each feature in each line are indicated by the vertical dashed lines. Lifetimes are indicated  in the top right corner of each plot. LT and LF stand for loop top and loop footpoint respectively. }
    \label{EIS intensity lineouts}
\end{figure}

The intensity evolution observed with the EIS slit over the region of interest including these two features is shown in Figure \ref{EISintensity}. In these EIS intensity maps, the flare loop top first appears as a strong localized intensity increase in the hot Fe {\scriptsize XXIV} 192.028 \AA\ and Fe {\scriptsize XXIII} 263.765 \AA\ lines and then as a c-shaped feature in the cooler lines. The loop footpoint is a fainter feature, which does not stand out in the intensity maps presented in Figure \ref{EISintensity}, but is clearly visible in the intensity profiles shown in Figure \ref{EIS intensity lineouts}. Figure \ref{EIS intensity lineouts} shows plots of the intensity evolution for the two features we are focusing our analysis on, the loop top and footpoint. To increase the signal, for the plots in Figure \ref{EIS intensity lineouts}, the EIS spectra were averaged over 7 y-pixels (y=255\as--262\as) and 2 exposures for the loop top and over 4 y-pixels (y=233\as--237\as) and 2 exposures for the loop footpoint. These ranges are also shown in Figures \ref{AIApanel} and \ref{EISintensity}. This binning resulted in a cadence of 32.4 s rather than 16.2 s for Figures \ref{EIS intensity lineouts} and \ref{Plasma parameters results}. The y-range selected for the loop top is slightly offset from the center of the c-shape feature to avoid the region around y=250\as where the Ca {\scriptsize XIV} 193.866 \AA\ line is affected by dust on the EIS detector \citep[see EIS Software Note No. 1; ][]{young_calibrating_2025}.

The intensity profiles shown in Figure \ref{EIS intensity lineouts} are then used to estimate the peak emission time and the emission lifetime in each line. For this, each emission evolution profile was fit with a Gaussian function. The peak emission time was defined as the Gaussian mean. The lifetime was defined as the duration over which the Gaussian function value was above 10\% of its amplitude, and is determined by the Gaussian mean and standard deviation. The measured lifetimes and peak emission times are summarized in Table \ref{Cooling_results}. 

The loop top is first observed in the Fe {\scriptsize XXIV} 192.028 \AA\ and Fe {\scriptsize XXIII} 263.765 \AA\ lines at 13:49 UT (peak emission; see Figure \ref{EIS intensity lineouts}). As time progresses, it becomes visible in the successively cooler lines, indicating that the loop top is cooling after the initial heating caused by the flare. Overall, the upper part of the flare loops cools down from $log(T/\mathrm{K})$ = 7.25 to $log(T/\mathrm{K})$ = 5.72 in approximately 18 minutes (time difference between the emission peaks in Fe {\scriptsize XXIV} 192.028 \AA\ and Fe {\scriptsize VIII} 194.661 \AA{}). The loop top starts to disappear from the Fe {\scriptsize VIII} 194.661 \AA\ line (the coolest line analysed here) after 14:07 (peak emission), indicating that the loop top has cooled below about $log(T/\mathrm{K})$ = 5.72. The loop top has a relatively short lifetime in Fe {\scriptsize XXIV} 192.028 \AA\ line (6 minutes), longer lifetimes in Ca {\scriptsize XIV} 193.866 \AA\, Ar {\scriptsize XIV} 194.401 \AA\ and Fe {\scriptsize XVI} 262.976 \AA\ (17--21 minutes) and similarly long lifetimes in Fe {\scriptsize XII} 195.119 \AA\ and Fe {\scriptsize VIII} 194.661 \AA\ (14--18 minutes), with the difference that the emission evolution in the latter is slightly double peaked, deviating from the clear Gaussian shape observed in the other lines. This could be indicative of two loop populations reaching this temperature at slightly different times.

The southern footpoint of the flare loops can also be observed as a fainter yet distinct feature. The loop footpoint is first observed at 13:49 UT in the Ca {\scriptsize XVII} 192.853 \AA\ and Fe {\scriptsize XVII} 269.420 \AA\ lines. It is not visible in the hotter Fe {\scriptsize XXIV} 192.028 \AA\ and Fe {\scriptsize XXIII} 263.765 \AA\ lines, indicating it does not get hot enough to emit at these temperatures. It is also not visible in the cooler Fe {\scriptsize XII} 195.119 \AA\ and Fe {\scriptsize VIII} 194.661 \AA\ lines. In Fe {\scriptsize XII} 195.119 \AA, there is strong emission in the footpoint region, possibly from a different feature along the field of view as it is relatively constant throughout the observation. This makes it difficult to distinguish the footpoint emission from the background emission. In Fe {\scriptsize VIII} 194.661 \AA, the footpoint emission is barely above the background emission, and does not show an obvious peak.

The loop top shows shorter lifetimes than the loop footpoint in almost all lines (with the exception of Ca {\scriptsize XVII} 192.853 \AA\ where they are equal). In addition, the peak emission times show a consistent delay of approximately 5 minutes between the loop top and the loop footpoint, indicating that the loop top is cooling through a given temperature approximately 5 minutes after the loop footpoint does.

\subsection{Evolution of Plasma Parameters}
The evolution of plasma temperature and composition in the flare loops was analysed using the intensity profiles in the loop top and footpoint locations highlighted in Figure \ref{EISintensity}.

\subsubsection{Plasma Density}
\label{S:Plasma Density}
First, the approximate electron density of the flare loops was estimated using the Ar {\scriptsize XIV} 194.401 \AA/187.962 \AA\ diagnostic. The Ar {\scriptsize XIV} 187.962 \AA\ line is blended with Fe {\scriptsize XXI} 187.929 \AA\ and Fe {\scriptsize IX} 187.950 \AA. The Fe {\scriptsize XXI} 187.929 \AA\ line cannot be deblended. It is a strong line which dominates the emission while Fe {\scriptsize XXIV} 192.028 \AA\ is strong (up until approximately 13:55 UT), but it can be neglected afterwards. The contribution of the Fe {\scriptsize IX} 187.950 \AA\ line to the observed Ar {\scriptsize XIV} 187.962 \AA\ intensity can be estimated using Fe {\scriptsize X} 184.537 \AA. A reference Hinode EIS quiet Sun spectrum (raster ID: 20140202\_141952) was used to calculate the Fe {\scriptsize X} 184.537 \AA/Fe {\scriptsize IX} 187.950 \AA\ ratio, which was found to be stable around a value of 25. So, to account for the Fe {\scriptsize IX} 187.950 \AA\ blend in the dataset presented here, the Fe {\scriptsize X} 184.537 \AA\ intensity is measured, divided by 25 and subtracted from the observed Ar {\scriptsize XIV} 187.962 \AA\ intensity before it is used for the density measurement. The density was found to be approximately $10^{10.50}~\mathrm{cm}^{-3}$ in both the loop top and footpoint.

\subsubsection{Plasma Temperature}
\label{S:Plasma Temperature}
The temperature evolution of the flare loop was estimated using a DEM analysis, assuming a Gaussian shaped function in linear space and the density value measured in Section \ref{S:Plasma Density}, at a few points along the flare evolution. This is a reasonable assumption as the flare loops are clearly cooling down: they brighten in a particular line for a few minutes, then disappear and brighten in a cooler line. Hence, at any one time, the plasma along the line of sight is mostly originating from a narrow temperature range. The spectra are averaged over multiple pixels to maximize the signal to noise ratio. For the loop top, the spectra are averaged over 7 pixels in the y direction (in between the dashed lines in Figure \ref{EISintensity}) and 7 pixels in the x direction, totaling 49 pixels. For the loop footpoint, the spectra are averaged over 4 pixels in the y direction (in between the dotted lines in Figure \ref{EISintensity}) and 9 pixels in the x direction, totaling 36 pixels. 

\begin{figure}[h]
    \centering
    \includegraphics[width=0.49\textwidth]{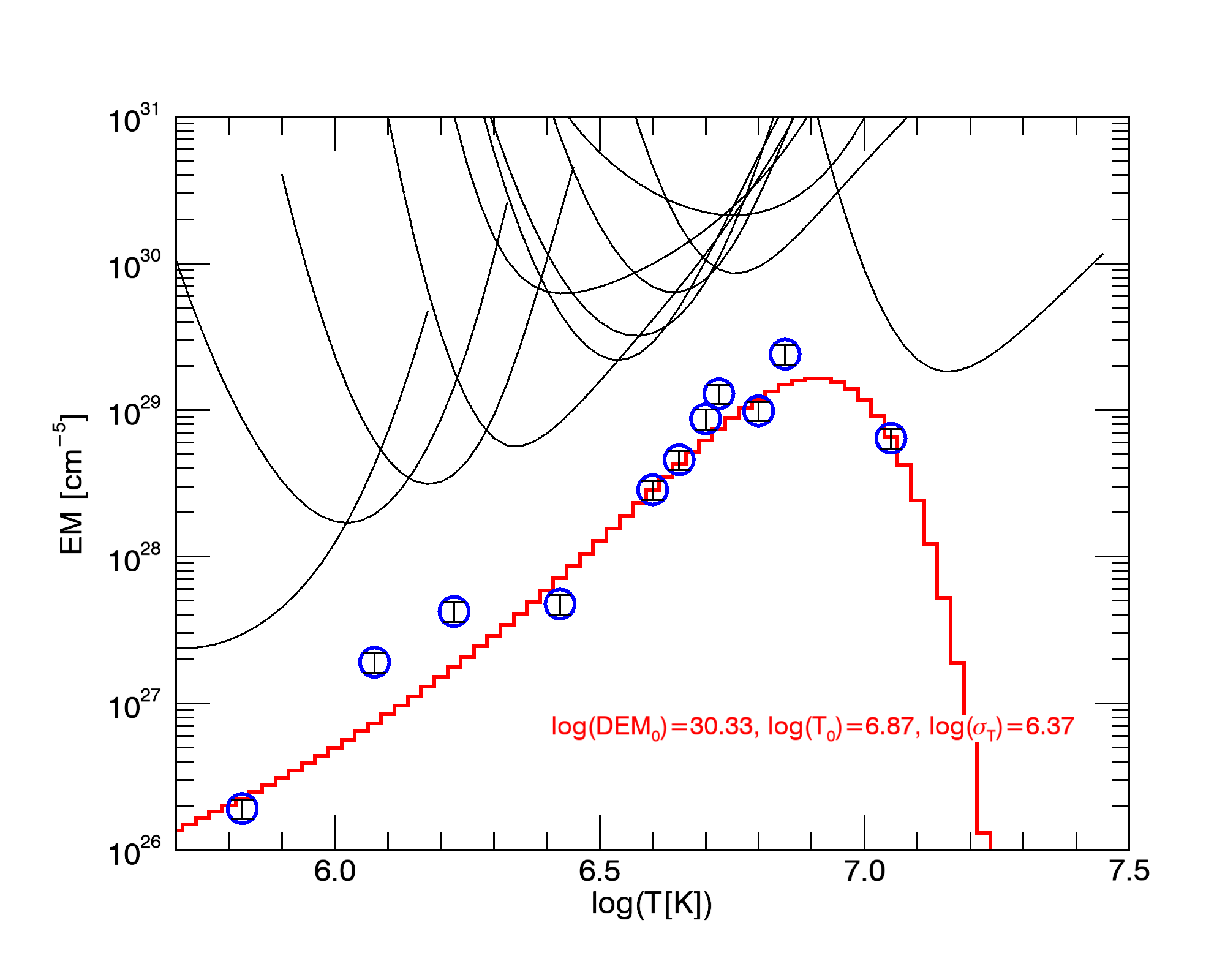}
    \caption{Example of EM curve derived from the DEM calculation for the third loop top data point (13:53-13:54 UT). Intensity data points are shown in blue circles.}
    \label{DEM plot example}
\end{figure}

The line list used as input for the DEM calculation is given in Table \ref{DEM intensities example}. The same line list was used for all the macropixels, with the exception that Fe {\scriptsize XXIII} 263.765 \AA{} was included in the calculations for the first 3 data points from the loop top (before 13:55 UT, where emission in strong in this line; see Figure \ref{EIS intensity lineouts}) but not for the later loop top data points or any of the loop footpoint ones (since these regions are not hot enough to emit in Fe {\scriptsize XXIII} 263.765 \AA{}). An example DEM, corresponding to one of the macropixels considered, is shown in Figure \ref{DEM plot example} and the corresponding parameters are listed in Table \ref{DEM intensities example}.

\begin{deluxetable}{cccccc}[t]
\centering
\tablecolumns{6}
\tablehead{
\colhead{Line} &
\colhead{$T_{\text{eff}}$} &
\colhead{$I_{\text{Obs}}$} &
\colhead{$\sigma_{I_{\text{Obs}}}$} &
\colhead{$I_{\text{Pred}}$} &
\colhead{$\frac{I_{\text{Obs}}-I_{\text{Pred}}}{\sigma_{I_{\text{Obs}}}}$}}
\tablecaption{DEM parameters corresponding to the example in Figure \ref{DEM plot example}, with units of K for $T_{\text{eff}}$, $\text{erg } \text{cm}^{-2}\text{s}^{-1}\text{sr}^{-2}$ for $I_{\text{Obs}}$, $\sigma_{I_{\text{Obs}}}$, $I_{\text{Pred}}$, and unitless for $\frac{I_{\text{Obs}}-I_{\text{Pred}}}{\sigma_{I_{\text{Obs}}}}$.}
\label{DEM intensities example}
\startdata
\shortstack{Fe {\scriptsize XXIII}\\263.765 \AA} & 7.05 & 513  & 79 & 522 & -0.1 \\
\shortstack{Ca {\scriptsize XVII}\\192.853 \AA}  & 6.80 & 12372 & 1856 & 14942 & -1.4 \\
\shortstack{Fe {\scriptsize XVII}\\269.420 \AA}  & 6.85 & 1301 & 196 & 803 & 2.5 \\
\shortstack{Ca {\scriptsize XV}\\182.863 \AA}    & 6.70 & 502 & 79 & 360 & 1.8 \\
\shortstack{Ca {\scriptsize XIV}\\193.866 \AA}   & 6.65 & 2180 & 327& 2024 & 0.5 \\
\shortstack{Ar {\scriptsize XIV}\\194.401 \AA}   & 6.60 & 371 & 56 & 371 & 0.0 \\
\shortstack{Fe {\scriptsize XVI}\\262.976 \AA}   & 6.73 & 14011 & 2102 & 8100 & 2.8 \\
\shortstack{Fe {\scriptsize XV}\\284.163 \AA}    & 6.43 & 28765 & 4316 & 43212 & -3.4 \\
\shortstack{Fe {\scriptsize XII}\\195.119 \AA}   & 6.23 & 5416 & 813 & 2260 & 3.9 \\
\shortstack{Fe {\scriptsize X}\\184.537 \AA}     & 6.08 & 1496 & 225 & 575 & 4.1 \\
\shortstack{Fe {\scriptsize VIII}\\194.661 \AA}  & 5.83 & 109 & 17 & 127 & -1.1 \\
\enddata
\end{deluxetable}

The temperature evolution is shown in Figure \ref{Plasma parameters results}. At the loop top, the plasma temperature cools down from $\log{T[K]}=6.95$ to $\log{T[K]}=6.61$ between 13:47 UT and 14:05 UT. The loop footpoint cools down from $\log{T[K]}=6.75$ to $\log{T[K]}=6.62$ between 13:44 UT and 14:04 UT. Overall, the loop top is heated to a higher temperature than the loop footpoint and is constantly hotter than the loop footpoint throughout most of the observation until they reach the same temperature towards the end of the observation.

\subsubsection{Plasma Composition}
\label{S:Plasma Composition}
To calculate the FIP bias, the Ca {\scriptsize XIV} 193.866 \AA\ (low FIP, FIP = 6.11 eV) and Ar {\scriptsize XIV} 194.401 \AA\ (high FIP, FIP = 15.76 eV) emission lines were used. The theoretical peak formation temperatures for the two lines are $\text{log}(\text{T}_{\text{MAX}}) = 6.58$ for Ca {\scriptsize XIV} 193.866 \AA\ and $\text{log}(\text{T}_{\text{MAX}}) = 6.53$ for Ar {\scriptsize XIV} 194.401 \AA{}. 

The DEM analysis was also used to derive the FIP bias of the flare loops. In the DEM analysis, Fe is used as the reference element as it contributes the highest number of lines. Hence, the DEM calculation provides the relative Ca/Fe and Ar/Fe abundances and these are then used to derive the relative Ca/Ar abundance. For Ca, the relative abundance of Ca to Fe is free to vary in the minimization process. This provides a Ca/Fe abundance ratio which represents a best fit for all the Ca lines considered \citep[the 'optimized' case, as described by][]{young_chianti_2025}. As there is only one Ar line (Ar {\scriptsize XIV} 194.401 \AA), the Ar abundance is not included in the minimization process, and calculated algebraically instead \citep[the 'algebraic' case, as described by][]{young_chianti_2025}.

The FIP bias errors are derived from the uncertainties associated with the measured intensities. Line fitting uncertainties are typically very small ($<$1\%), however, they represent only one component of the overall uncertainties. Estimating uncertainties associated with the spectrometer's radiometric calibration or with the atomic data is difficult. To broadly capture these additional sources of uncertainty, it is standard practice to add a blanket uncertainty \citep[typically 15-20\%;][]{young_chianti_2025} to the intensity of each measured line intensity (and added in quadrature to the line fitting  uncertainty). Pre-flight laboratory measurements suggest a radiometric calibration uncertainty of 22.7\% \citep{lang_laboratory_2006}, while the most recent in-flight radiometric calibration \citep{del_zanna_hinode_2025} suggests 20\%. The relative calibration, however, which is important for the FIP bias calculation, is better than 20\%, hence we use a values of 15\% in the analysis presented here. For further details on the uncertainties and references to earlier work, see also the discussion of \citet{kucera_spectroscopic_2019}.

\begin{figure}
    \centering
    \includegraphics[width=0.49\textwidth]{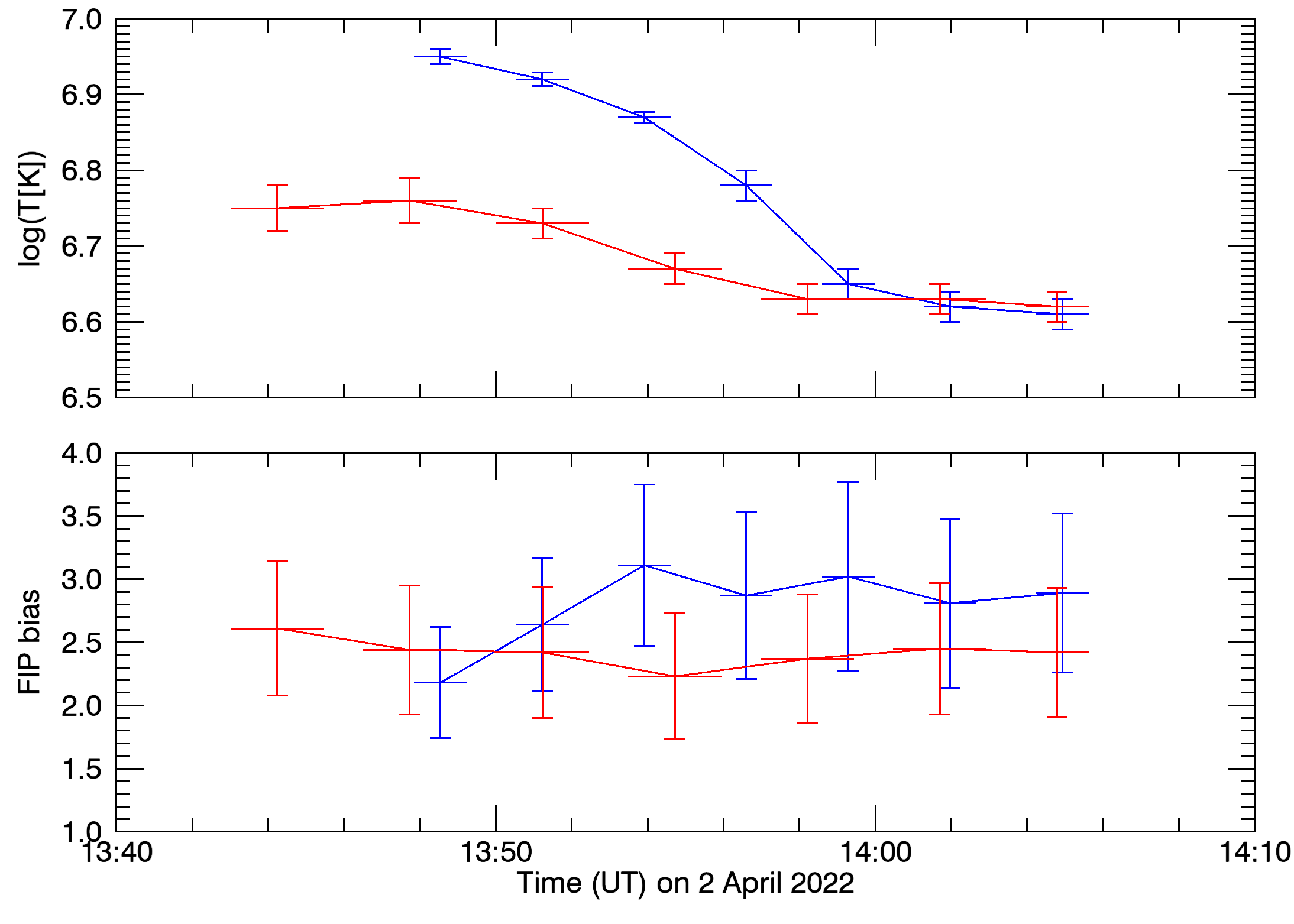}
    \caption{Evolution of Gaussian DEM centroid (top) and associated FIP bias (bottom) for the loop top (blue) and loop footpoint (red).}
    \label{Plasma parameters results}
\end{figure}

The FIP bias evolution is shown in Figure \ref{Plasma parameters results}. Both the loop top and footpoint FIP bias are relatively constant throughout the observation. The average FIP bias at the loop top, $2.8 \pm 0.2$, is higher than at the loop footpoint, $2.4 \pm 0.2$. This could provide an explanation as to why the lifetimes are generally higher for the loop top than the loop footpoint, as plasma composition influences the radiative cooling process of the flare loops.

\section{Radiative Cooling Modelling}
\subsection{The Radiative Loss Function}
\begin{figure}
    \centering
    \includegraphics[width=0.5\textwidth]{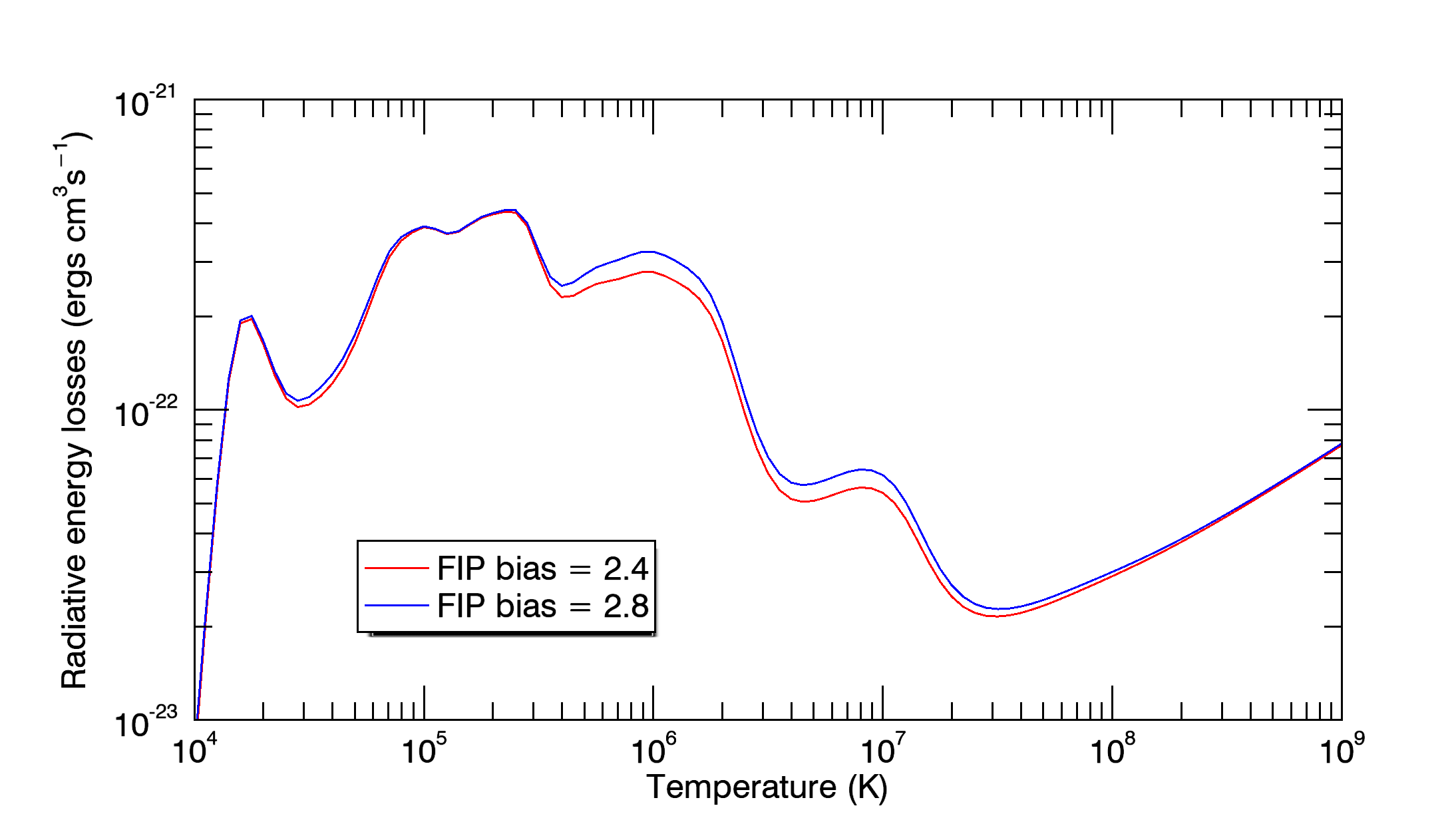}
    \caption{Radiative energy losses, $Q(T, N_e)$, as a function of electron temperature in the case of plasma with FIP bias = 2.4 (red) and FIP bias = 2.8 (blue), assuming the default CHIANTI ionisation file and an electron density of $10^{10} \mathrm{cm}^{-3}$.}
    \label{Radiative Loss Function}
\end{figure}

The radiative loss, i.e., total radiated power emitted from an optically thin plasma, is given by:
\begin{equation}
    L_r = N_e N_H Q(T, N_e),
\end{equation}
where $N_e$ and $N_H$ are the electron and hydrogen densities, respectively, $T$ is the electron temperature and $Q(T, N_e)$ is the radiative loss function. The main contributions to the radiative loss function are bound--bound line emission and continuum radiation, i.e., free--free emission and free--bound radiative recombination. In the temperature range $log(T/\mathrm{K})=4.5-7.0$, bound--bound emission from abundant elements such as C, O, Si and Fe dominates, with O having the highest contribution around $log(T/\mathrm{K}) \approx 5.3$ and Fe around $log(T/\mathrm{K}) \approx 6.0$. For $log(T/\mathrm{K})\gtrsim 7.0$, however, continuum processes dominate \citep[see e.g.][]{aschwanden_physics_2004}. The radiative loss function depends on the plasma composition of the radiating plasma, particularly in the temperature range where it is dominated by emission from low-FIP elements \citep{cook_effect_1989}. Figure \ref{Radiative Loss Function} exemplifies this, showing the variability of the radiative loss function with temperature for plasma with FIP bias = 2.8 and FIP bias = 2.4, as calculated by CHIANTI using the rad\_loss routine available in IDL SolarSoft. The selected FIP bias values represent the average FIP bias in the loop top and loop footpoint features respectively. The input abundances for the radiative loss calculations make use of the photospheric abundances of \citet[][sun\_photospheric\_2015\_scott.abund file in the SolarSoft database]{scott_elemental_2015} as a starting point and abundances of low-FIP elements are increased by the FIP bias in each case. There are significant differences between the two functions around $log(T/\mathrm{K}) \approx 5.3-7.0$, where emission from Fe (low-FIP) lines dominates the radiative losses, but the differences are much weaker in the temperature ranges where emission from O (high-FIP) lines or continuum emission dominate.

\subsection{EBTEL Simulations}
The effect of varying plasma composition in the cooling of flare loops was studied with a simulation using the enthalpy-based thermal evolution of loops (EBTEL) model \citep{klimchuk_highly_2008, cargill_enthalpy-based_2012, cargill_enthalpy-based_2012-1}. EBTEL is a 0D hydrodynamic model which simulates the radiative cooling process in a loop in response to a heating event, predicting the evolution of average loop parameters (e.g., temperature, density, pressure) with time. Using these loop parameters, synthetic line emission can be derived and compared to observations. 

\begin{figure}
    \centering
    \includegraphics[width=0.46\textwidth]{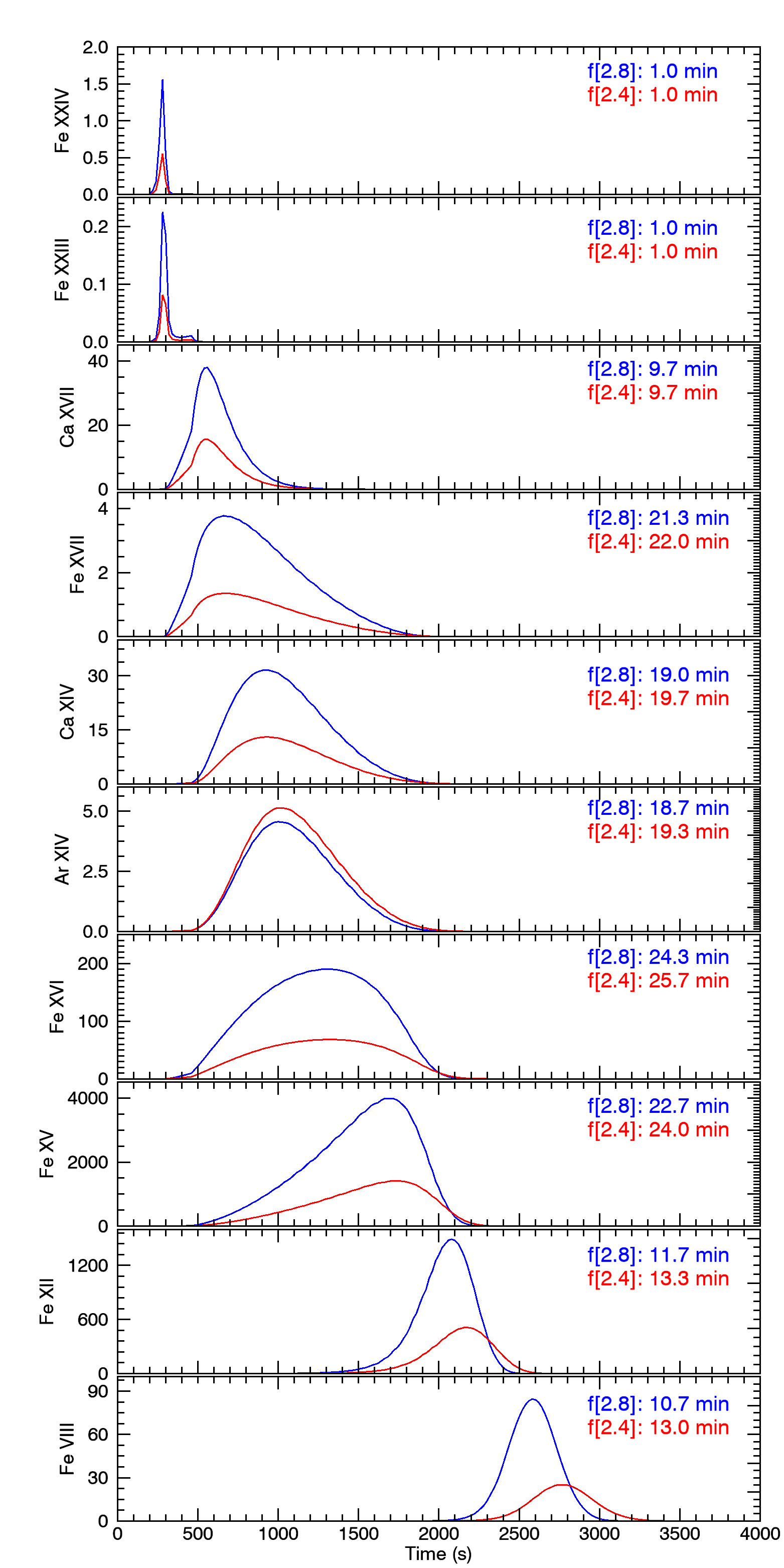}
    \caption{Synthetic intensity evolution of the emission lines analysed in Figure \ref{EIS intensity lineouts} computed using an EBTEL simulation of a heating event in a single loop strand, assuming a plasma with a FIP bias = 2.8 (blue) and a FIP bias = 2.4 (red) respectively. The lifetime of a given line in each plasma composition case is provided on the right hand side. }
    \label{EBTELsimulations}
\end{figure}

For the simulation presented here, we use a single strong heating event of $0.023 \text{ erg} \text{ cm}^{-3} \text{s}^{-1}$ with a step--function starting 200 s and stopping 450 s after the simulation began. A constant background heating rate of $5 \times 10^{-6} \text{ erg} \text{ cm}^{-3} \text{s}^{-1}$ was assumed. The heating function was chosen such that the simulated temperature of the loop plasma matched the observations and the simulated density gets as close to the observed density as possible. However, the density is a little too high for EBTEL: for our chosen parameters, we can reach approximately $10^{9.50}~\mathrm{cm}^{-3}$ before the simulation collapses, so we adopted this value. The loop is also assumed to be a single strand. As the loop cools down, emission increases in progressively cooler lines. Figure \ref{EBTELsimulations} shows the simulated intensity evolution for the Hinode EIS lines analysed. Lifetimes for each of the lines are defined as the time spent above 10\% of the maximum intensity of the line, the same as the lifetimes of the observed EIS lines to facilitate comparison. 

Simulations show that, in the FIP bias = 2.8 case, line intensities are higher and line lifetimes are shorter, i.e., the loop cools down faster. Conversely, in the FIP bias = 2.4 case, line intensities are lower and line lifetimes are longer, i.e., the loop cools down more slowly. The difference in line lifetimes is barely noticeable for the very hot lines (Fe {\scriptsize XXIV} 192.028 \AA\ here), but increases for cooler lines. This can likely be explained by the role of Fe in modulating the radiative cooling function. At higher temperatures ($log(T/\mathrm{K}) \gtrsim 7.0$), the continuum emission dominating the radiative cooling function is not dependent on plasma composition, so the lifetimes in the two composition cases are very similar. As the loop cools down, bound--bound emission, mainly from Fe lines, begins to dominate. As Fe is a low-FIP element, it will be overabundant in the FIP bias = 2.8 case which increases the emission and also increases the cooling rate which speeds up the cooling process. Going from hotter to cooler lines, the differences between lifetimes in the FIP bias = 2.8 and FIP bias = 2.4 cases increases and so does the gap between the peak emission times. 

\section{Comparison between Hinode EIS Observations and EBTEL Simulations}
\begin{deluxetable}{ccccc}[t]
\centering
\tablecolumns{5}
\tablehead{
\colhead{Line} &
\multicolumn{2}{c}{EIS Observations} &
\multicolumn{2}{c}{EBTEL Simulations} \\
\cline{2-3} \cline{4-5}
\colhead{} &
\colhead{LT} &
\colhead{LF} &
\colhead{LT} &
\colhead{LF}}
\tablecaption{Summary of cooling lifetimes from Hinode EIS Observations and EBTEL simulations. LT and LF stand for loop top and loop footpoint, respectively.}
\label{Cooling_results}
\startdata
\shortstack{Fe {\scriptsize XXIV}\\192.028 \AA}  &  5.8 min  &     -    &  1.0 min &  1.0 min \\
\shortstack{Fe {\scriptsize XXIII}\\263.765 \AA} &  6.4 min  &     -    &  1.0 min &  1.0 min \\
\shortstack{Ca {\scriptsize XVII}\\192.853 \AA}  & 20.7 min  & 21.4 min &  9.7 min &  9.7 min \\
\shortstack{Fe {\scriptsize XVII}\\269.420 \AA}  & 22.2 min  & 25.3 min & 21.3 min & 22.0 min \\
\shortstack{Ca {\scriptsize XIV}\\193.866 \AA}   & 17.2 min & 28.6 min & 19.0 min & 19.7 min \\
\shortstack{Ar {\scriptsize XIV}\\194.401 \AA}   & 18.8 min & 30.7 min & 18.7 min & 19.3 min \\
\shortstack{Fe {\scriptsize XVI}\\262.976 \AA}   & 20.6 min & 27.0 min & 24.3 min & 25.7 min \\
\shortstack{Fe {\scriptsize XV}\\284.163 \AA}    & 18.8 min & 22.4 min & 22.7 min & 24.0 min \\
\shortstack{Fe {\scriptsize XII}\\195.119 \AA}   & 14.7 min &    -   & 11.7 min & 13.3 min \\
\shortstack{Fe {\scriptsize VIII}\\194.661 \AA}  & 17.2 min &    -   & 10.7 min & 13.0 min \\
\enddata
\end{deluxetable}

Comparing the EBTEL predictions with the EIS observations,  two main similarities are observed. First, EBTEL simulations show shorter (longer) lifetimes in the higher (lower) FIP bias case. Similarly, EIS observations show shorter (longer) lifetimes in the loop apex (footpoint) which has higher (lower) FIP bias. The average measured FIP bias is  larger in the loop apex than the loop footpoint, indicating that the shorter lifetimes observed here could be a result of plasma composition influencing the flare loop radiative cooling.

Second, EBTEL simulations show higher (lower) intensities in the high (low) FIP bias case, similar to how EIS observations indicate higher (lower) intensities in the loop apex (footpoint) which has higher (lower) FIP bias. This is not related to the radiative cooling process itself, but rather a result of the intensity being proportional to the abundance of the emitting element.

Finally, EBTEL predicts that the difference in peak emission time for the higher and lower FIP bias cases increases going towards cooler lines. This is not observed in the EIS data, where the difference in peak emission time for the loop apex and footpoint is consistently around 5 minutes for all the lines in the study (see vertical dashed lines in Figure \ref{EIS intensity lineouts}). It could be that an obvious difference is not detected because the peak times for the loop footpoint are not observed in the Fe {\scriptsize XII} 195.119 \AA\ and Fe {\scriptsize VIII} 194.661 \AA\ lines, which is where this difference is most visible in the EBTEL simulations. 

One limitation of the simulations is that they assume fixed abundances, i.e. the simulations were run for FIP bias values of 2.8 and 2.4. In the observations, however, the FIP bias shows slight variations around these values (within the uncertainties). Such variation is expected to also affect the cooling times, although to a lower degree. Recent work by \citet{reep_modeling_2024, benavitz_spatiotemporal_2025} investigated the role of time-varying FIP bias in changing the cooling rate and, therefore, cooling times in flare loops. In the present work, however, we are interested in the qualitative effects, so the two cases we selected are sufficient. 

It is important to also note that a simple model is used here, assuming a single step function-like heating event on a single strand. In reality, the heating function generated by a flare is likely to be more similar to a succession of smaller energy releases as more and more field lines are pushed into the reconnection region and reconnect. And, each loop might be composed of more than one strand \citep{brooks_solar_2012}. In addition, the model is 0D, so it does not allow for variations in loop properties (e.g. composition, temperature, density, etc.) along the loop. However, the simulations qualitatively show the effect of varying plasma composition in speeding up or slowing down the radiative cooling process and they also provide approximate cooling time differences for the two FIP bias values considered. Previous studies suggest that loop geometry likely plays an important role in flare loop evolution as well \citep{reep_geometric_2022} and this was not investigated in the present work.

Finally, the IDL EBTEL implementation used for the simulations only solves the single fluid equations. The C++ implementation, ebtel++, is able to treat the electron and ion populations separately. This might yield slightly different results for plasma above roughly $log(T/\mathrm{K})=6.7$, i.e., affecting the simulated Fe {\scriptsize XXIV} 192.028 \AA\ to Fe {\scriptsize XVII} 269.420 \AA. Below  $log(T/\mathrm{K})=6.7$, however, i.e., where the observed trends are the strongest, there is little or no difference between the two-fluid and single-fluid models. 

\section{Summary and Discussion}
This work investigates the evolution of plasma composition in flare loops as well as its link to the radiative cooling process of the flare loops. 

\subsection{Plasma Composition Observations}
The EIS spectroscopic observations indicate a clear difference in the FIP bias at the loop top and footpoint. Results show a constant FIP bias of approximately $2.4 \pm 0.2$ in the loop footpoint and a stronger FIP bias of $2.8 \pm 0.2$ in the loop apex. Previous flare observations using the same Ca {\scriptsize XIV} 193.886 \AA{}/Ar {\scriptsize XIV} 194.401 \AA\ diagnostic found FIP bias values of approximately 1 and 2 in two separate flares observed by Skylab \citep{young_arca_1997} and 3-4 in a small brightening \citep{to_evolution_2021}. Interestingly, all these values observed in flares are lower than the FIP bias measurements of approximately 3 to 8 obtained with the same diagnostic in an active region \citep{mihailescu_intriguing_2023}. 

The relatively high FIP bias values measured in both locations may appear in contradiction with previous work that found low and decreasing FIP bias after the peak in the flare soft X-ray emission \cite[e.g.,][]{del_zanna_spectral_2013, warren_absolute_2014} or very low FIP bias values \citep{doschek_photospheric_2018} and even the inverse FIP effect \citep{doschek_anomalous_2015, doschek_mysterious_2016, doschek_sunspots_2017, baker_transient_2019} associated with flare loop footpoints, all indicating chromospheric evaporation transporting plasma from below the fractionation region into the corona. However, a few factors could provide an explanation for this apparent disagreement.

First, the temperature of the emission used for the composition analysis. Previous full-disk studies have analysed high temperature emission \citep[$\sim$10 MK; e.g.][]{phillips_solar_2012, del_zanna_spectral_2013, warren_absolute_2014, sylwester_solar_2014, dennis_solar_2015, mondal_evolution_2021, nama_coronal_2023}, while in this study we analyse emission at $\sim$4 MK. One could imagine a scenario where the observed FIP bias depends on the phase of the flare and the timing of the measurement relative to the heating event. In such a scenario, our observations likely sample the flare loop plasma after it had begun to cool from the initial heating episode, while previous studies may have captured the emission closer to the peak of the heating stage. The plasma composition could evolve during this interval, meaning that the different observations probe distinct stages of the flare evolution and its associated composition. During the early phase, the more intense energy release and heating could lead to a lower FIP bias, while, in the later stages, the decrease in energy release and heating are likely to lead to a higher FIP bias. Our higher FIP bias measurements during the later flare stages would be consistent with this scenario.

Second, the amount of flare energy released and flare transport mechanism, as well as their spatial and temporal variations. If the flare energy is deposited higher up in the chromosphere, i.e. above the fractionation region, then the resulting chromospheric evaporation brings plasma with higher FIP bias in the corona, consistent with our observations. The height of the energy deposition depends on both the energy transport mechanism \citep[simulations suggest that electron beams deposit their energy in a more shallow chromospheric layer than Alfvén waves;][]{allred_unified_2015, kerr_simulations_2016} and the energy budget (a lower amount of energy will lead to a lower penetration depth). Recent work by \citet{kerr_spatial_2026} showed that transport mechanisms can vary spatially in flares and one could also expect that the amount of energy released varies with time during the flare evolution as well, tying into the point made above regarding the timing of the observed emission. While the event is classed as M3.9, the flare loop arcade studied here is somewhat isolated from the main flare loops, so it seems reasonable to assume a lower amount of energy might be released into this arcade.

Third, the FIP bias diagnostic used and calculation method. Studies using hot emission typically measure absolute abundances from line intensities relative to continuum emission, while our measurement calculates relative abundances using intensity line ratios. In addition, previous studies showed that FIP bias measurements from different element diagnostics might not necessarily agree \citep[e.g.][]{mihailescu_intriguing_2023}. This could be significant in flares as \citet{dennis_solar_2015} suggested the separation between low-FIP and high-FIP elements might be closer to 7 eV in flares, as opposed to the typically agreed value of 10 eV. If this is indeed the case, then elements typically considered low-FIP, such as Fe or Si, would actually become high-FIP elements in flares.

Therefore, what might initially appear as disagreeing results could, in fact, be a set of complementary constraints that provide insight into plasma evolution in flares and its tight relationship to flare dynamics. Finally, while reports of low FIP bias values associated with flares have dominated the literature \citep[e.g.][]{del_zanna_spectral_2013, warren_absolute_2014, sylwester_solar_2014, mondal_evolution_2021, nama_coronal_2023}, high FIP bias values have been also been observed \citep[e.g.][]{phillips_solar_2012, dennis_solar_2015}. This strengthens the point that the plasma composition behavior observed likely depends on the particular characteristics of the studied events, the temperature and timing of the emission studied and also on the methods and diagnostics used to estimate the FIP bias values.

The higher FIP bias at the loop top is consistent with the reconnection outflow idea proposed by \citet{to_spatially_2024}. This reconnection downflow was proposed to bring plasma from the reconnection region to the top of the observed loops. Since the reconnecting loops are part of the pre-flare active region loops, it is reasonable to assume they are filled with plasma with high FIP bias hence driving the FIP bias increase seen at the flare loop top. 

\subsection{Flare Loop Cooling}
The Hinode EIS observations show consistently shorter cooling times, measured as the emission lifetime in a series of lines, at the loop top than the loop footpoint. The EBTEL simulations show that an increased FIP bias would lead to a faster radiative cooling rate and, therefore, shorter cooling times. This indicates that the difference in cooling rates observed by Hinode EIS is due to the difference in plasma composition between the two features. It would then be reasonable to speculate that a larger difference in cooling rates might be observed for flares where the FIP bias gradient along the flare loop is higher. Such an example would be the X-class flare studied by \citet{doschek_photospheric_2018} and \citet{to_spatially_2024}. Further work is required to explore whether this is indeed the case and, if yes, what effects this might have on the flare dynamics and heat transfer along the loop. The role of conduction, which might also be important in cooling the loop and, therefore, affect temperature evolution along the loop is not considered in this work.

\begin{acknowledgments}
We thank the anonymous referee for their insightful suggestions which helped improve the manuscript. T.M. acknowledges funding from STFC PhD studentship ST/V507155/1, PHaSER 80NSSC21M0180 grant and Solar Orbiter SPICE funding at the NASA Goddard Space Flight Center. T.M. is also grateful for a Royal Astronomical Society E. A. Milne travelling fellowship to carry out part of this work. The work of D.H.B. was performed under contract to the Naval Research Laboratory and was funded by the NASA Hinode program. D.B. is funded under Solar Orbiter EUI Operations grant number ST/X002012/1 and Hinode Ops Continuation 2022-25 grant number ST/X002063/1. Hinode is a Japanese mission developed and launched by ISAS/JAXA, collaborating with NAOJ as a domestic partner, and NASA and STFC (UK) as international partners. Scientific operation of Hinode is performed by the Hinode science team organized at ISAS/ JAXA. Support for the post-launch operation is provided by JAXA and NAOJ (Japan), STFC (UK), NASA, ESA, and NSC (Norway). CHIANTI is a collaborative project involving University of Cambridge (UK), NASA Goddard Space Flight Center (USA), the University of Michigan (USA) and George Mason University (USA).
\end{acknowledgments}

%

\vspace{5mm}





\bibliography{references}{}
\bibliographystyle{aasjournal}



\end{document}